\documentclass[manuscript]{acmart}
\usepackage{nidanfloat}
\usepackage{placeins}
\usepackage{multirow}

\AtBeginDocument{%
  \providecommand\BibTeX{{%
    \normalfont B\kern-0.5em{\scshape i\kern-0.25em b}\kern-0.8em\TeX}}}

\copyrightyear{2022} 
\acmYear{2022} 
\setcopyright{acmlicensed}
\acmConference[CIKM '22]{Proceedings of the 31st ACM International Conference on Information and Knowledge Management}{October 17--21, 2022}{Atlanta, GA, USA}
\acmBooktitle{Proceedings of the 31st ACM International Conference on Information and Knowledge Management (CIKM '22), October 17--21, 2022, Atlanta, GA, USA}
\acmPrice{15.00}
\acmDOI{10.1145/3511808.3557087}
\acmISBN{978-1-4503-9236-5/22/10}

\begin{document}

\title{
Ensure A/B Test Quality at Scale with Automated Randomization Validation and Sample Ratio Mismatch Detection}

\author{Keyu Nie}
\authornote{Both authors contributed equally to this research.}
\email{keyunie@gmail.com}
\author{Zezhong Zhang}
\authornotemark[1]
\email{zezzhang@ebay.com}
\affiliation{%
  \institution{eBay, Inc.} 
  \city{San Jose}
  \state{California}
  \country{USA}}

\author{Bingquan Xu}
\email{bingquanx@gmail.com}
\affiliation{%
  \institution{eBay, Inc.} 
  \city{Shanghai}
  \country{China}}

\author{Tao Yuan}
\email{teyuan@ebay.com}
\affiliation{%
  \institution{eBay, Inc.} 
  \city{San Jose}
  \state{California}
  \country{USA}}

\renewcommand{\shorttitle}{Running High Quality Experiments at Scale}

\begin{abstract}
 eBay's experimentation platform runs hundreds of A/B tests on any given day. The platform integrates with the tracking infrastructure and customer experience servers, provides the sampling service for experiments, and has the responsibility to monitor the progress of each A/B test. There are many challenges especially when it is required to ensure experiment quality at the large scale. We discuss two automated test quality monitoring processes and methodologies, namely randomization validation using population stability index (PSI) and sample ratio mismatch (a.k.a. sample delta) detection using sequential analysis. The automated processes assist the experimentation platform to run high quality and trustworthy tests not only effectively on a large scale, but also efficiently by minimizing false positive monitoring alarms to experimenters.
\end{abstract}




\maketitle

\section{Introduction}

A trustworthy experimentation platform is the foundation to support eBay's data driven business and product decision making. A/B tests running on the platform are based on the randomized controlled trial (RCT) \cite{kohavi2009controlled, kohavi2020trustworthy} that is widely regarded as the gold standard of causal inference. However, in the industry, such as Google \cite{kohavi2020trustworthy}, Meta, Microsoft \cite{kohavi2009controlled,kohavi2020trustworthy,microsoftsrm}, Linkedin \cite{linkedin_chen2019b}, Twitter, Yahoo \cite{yahoosrm} and others, studies with randomization quality related concerns are commonly reported. These quality related issues are flawed and harmful because they can lead to incorrect analysis and conclusion. 

In practice, there are two types of randomization issues that draw our attention. The first is imperfect randomization, which occurs during the traffic/sample randomization assignment. Consider a platform provides a randomization engine that 
divides web traffic into small buckets evenly (say 100 buckets, so each bucket represents 1\% out of the total traffic). If one of the bucket had more than 1\% traffic (for example, 1.1\% actual traffic or 10\% relative deviation), we would suspect a problem with our randomization engine. As a result, a sample randomization validation check is critical to ensure that samples are dispersed evenly in a uniform Multinomial distribution. 
In fact, "unhappy randomization" and operating issues are the most common causes of faulty randomization results \cite{morgan2012rerandomization}.

The other common A/B test quality issue arises from mis-matching of the expected sample ratio to the observed sample ratio between test and control groups in triggered users. For example, consider an experiment with $10,000$ users total traffic in which each user has an equal chance of experiencing test and control features, a.k.a. 50-50 split. The expected sample ratio would be 50/50=1. The observed sample ratio is 2/3 if the experiment ends with just $2,108$ (test) and $3,183$ (control) triggered users in the test and control groups respectively (similar example can be found in \cite{optimizelySRMs,optimizelyblog}). The sample ratio mismatch between the expected and observed could also raise concerns on the validity of an experiment. It may indicate issues in the experiment implementation (as Microsoft suggests in \cite{microsoftsrm}), e.g., errors in the event tracking, user experience service programming code, and bot filtering-related data processing. 

These errors are usually beyond the control of experimentation platform. However it is the role of the experimentation platform to monitor and send accurate alerts so that experimenters can investigate on the ramification of the situation as soon as possible. While randomization validation and sample ratio mismatch detection are used to monitor the quality of A/B tests, the majority of these alerting approaches themselves are statistical hypothesis testings. The main challenge of the alerting is to balance between the essential need of early detection of quality issues (sensitivity) and the minimization of false alarms (due to the nature of randomness). 

Standard goodness-of-fit tests, such as  Pearson ${\chi}^2$ test \cite{chisq_cochran1952chi2,simard2011computing}, Kolmogorov–Smirnov test (KS test) \cite{massey1951kolmogorov} and Anderson-Darling test (AD test) \cite{scholz1987_ad_test,razali2011power} provided in Google, Microsoft and Linkedin \cite{kohavi2020trustworthy}, are common solutions for randomization validation of uniformly Multinomial distributions.
According to Yahoo \cite{yahoosrm}, Linkedin \cite{linkedin_chen2019b} and Microsoft \cite{microsoftsrm}, the Pearson ${\chi}^2$ test and two sample t-test are utilized in sample ratio mismatch detection. Our initial version of sample ratio mismatch detection algorithm was based on a t-test that was overly sensitive and generated tons of noisy signals. For example, the observed sample ratio of 2/3 might be OK at the early stage with only 2 and 3 users in each group, but it would worry us if the sample size was extended to 2000 and 3000 users, respectively. Our current algorithm is built on a series of adaptive sequential tests that significantly reduces false alarms while maintaining the sensitivity we need to effectively monitor sample delta for hundreds of experiments daily without spamming experiment owners.


In the paper, we describe two levels of test quality monitoring and alerting, (1) sample randomization check for traffic assignment, and (2) sample ratio mismatch detection for the triggered (or qualified) samples. Both are targeted to minimize false positives in their processes while attaining good recall of the signals. The methodologies involve a novel Population Stability Index (PSI \citep{psi_yurdakul2019statistical}) based test and a sequential probability ratio test (SPRT \citep{waldseq,microsoftseq,yandexseq,linkedinseq,optimizely,zhao2018safely}). To our knowledge, we are the first to automate the methods (in early 2019) in a large scale experimentation platform to continuously monitor experiment quality.

The paper is structured as follows: first we briefly describe our randomization engine in section \ref{Architecture}, then we introduce the PSI test for randomization check in section \ref{psi}, it follows with our sequential probability ratio test for sample delta detection in section \ref{sprt}. We provide the application use cases of diagnosing test issues using the methodologies. We conclude in section \ref{smry}.

\section{Randomization Architecture}\label{Architecture}

\subsection{Sample Assignment}

In the experimentation platform, the randomization engine leverages the standard MD5 algorithm and math modulo (MOD) function to hash samples uniformly 
into small buckets.
\begin{equation}
\label{eq_hash}
\begin{aligned}
  \big{(}\mathbf{Hash}(x, seed) \mod B \big{)} \sim \mathbf{Multinom}(\frac{1}{B})
\end{aligned}
\end{equation}
where $x$ is the sample identifier (e.g., cookie id), $\mod$ is the arithmetic modulo operator. $\mathbf{Hash}(x, seed)$ is the hashing function with a randomization seed (e.g., plane-id). $B$ is the total number of traffic buckets (e.g., 100).

\begin{figure}[!hb]
  \centering
  \includegraphics[width=\linewidth]{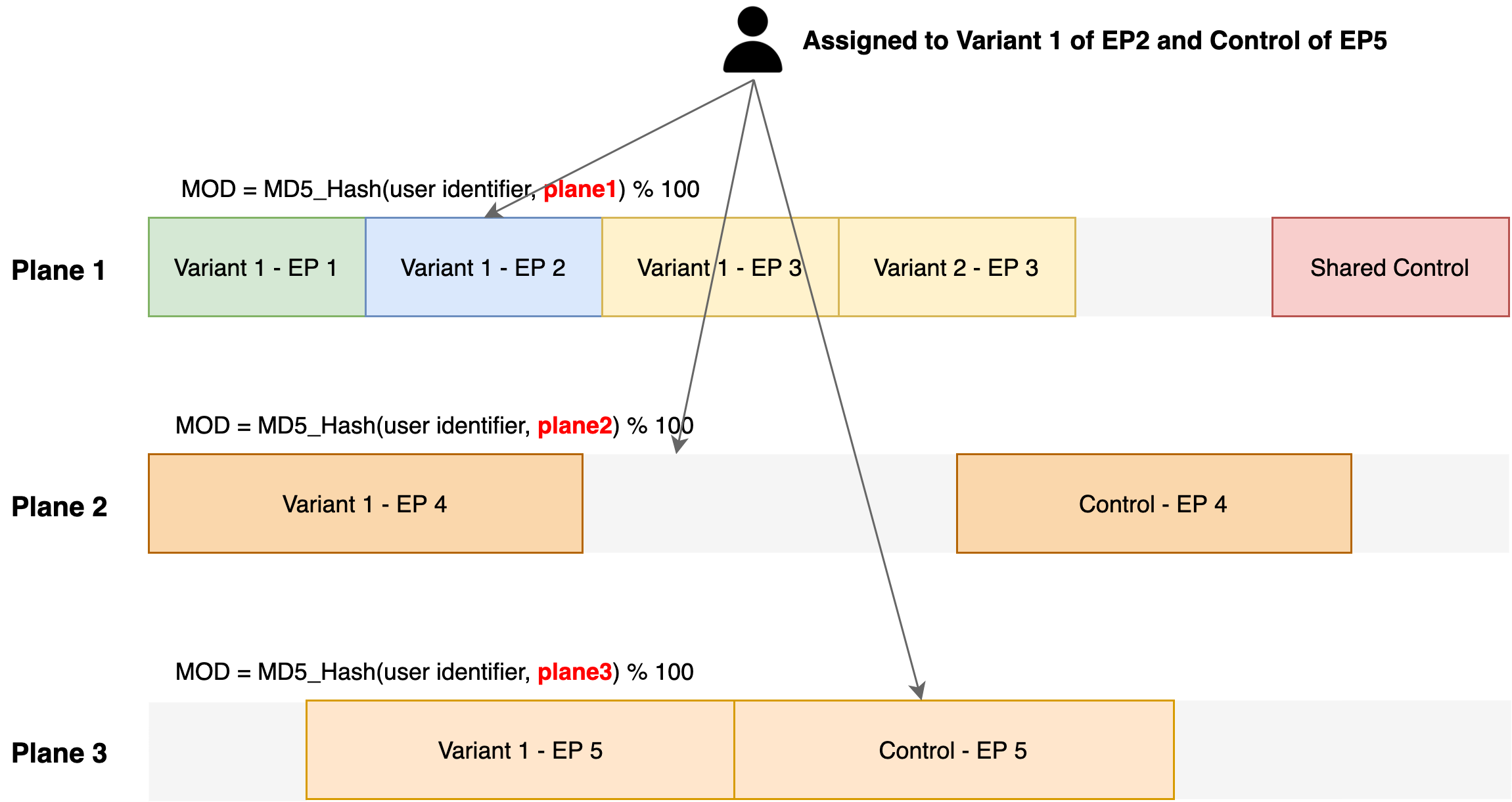}
  \caption{Exclusive and Orthogonal Tests (Each plane id is associated with a distinct random value.)}
  \label{fig:orth_test}
  \vspace{-1.0em}
\end{figure}

The experimentation platform supports exclusive and orthogonal testing modes \footnote{Independent features can be tested orthgonally by using different "planes", e.g. one plane for ranking algorithm, another for UI presentation. However, if two features can cause severe collision of user experience, they must be tested exclusively by using a shared "plane".}, see Figure \ref{fig:orth_test}. The exclusive mode allows multiple experiments running on a shared plane (by hashing with the same randomization seed). The orthogonal mode runs an independent feature experiment on its own plane (by hashing with its own randomization seed). Each plane covers the whole online traffic, and is divided into 100 uniform buckets. Buckets can be grouped into "swim lanes" (e.g., buckets with mod value from 0-49) and assigned to test and control variants. 
A bucket on a plane can only be assigned to one variant group of an experiment at a given time. 
The more buckets a variant occupies, the more traffic it is expected to have. 

For an experiment, one user is assigned a mod value that corresponds to a specific bucket on a certain plane. This mod value varies from different planes. The randomization engine should distribute the user traffic uniformly and randomly into test and control buckets for an experiment. A randomization failure can cause unbalanced assignments of the samples among the buckets. When this happens, the experiment becomes untrustworthy and is extremely difficult to rectify using post-processing techniques.

\clearpage


\subsection{Sample Triggering}

\begin{figure}[!hb]
  \includegraphics[width=0.9\linewidth]{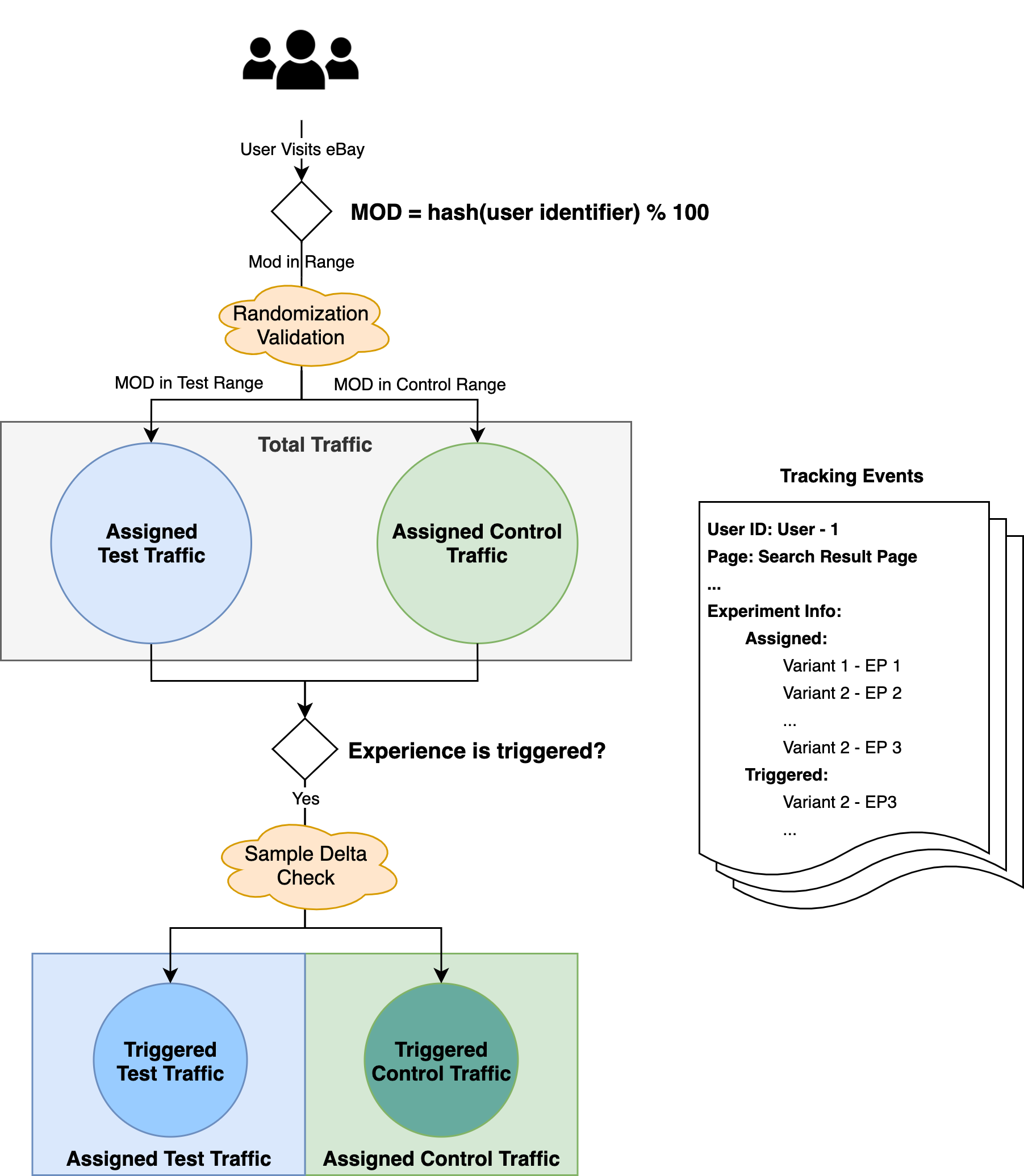}
  \caption{Experiment Sampling Process}
  \label{fig:srm}
  \vspace{-1.0em}
\end{figure}

After a sample user is assigned to a variant in an experiment by the randomization engine, the user is only considered triggered, when he or she actually visits the eBay application, see Figure \ref{fig:srm}.

\clearpage

Failure can occur in the sample assignment and qualification stages. A randomization failure can manifest itself in a variety of ways. Some of them are simple and straightforward to spot. For example, allocating the same user to both the test and the control is incorrect yet easy to detect. However, sample ratio mismatch, or the discrepancy between realized and expected traffic ratio between groups, can be difficult to investigate.
Detecting such issues as early as possible can help experimenters to fix the underlying issues without further loss of time and resources. Such detection functions are essential for the effectiveness of running experiments at scale and ensures the trustworthiness of all A/B tests.

\section{Randomization Validation with a Population Goodness-of-Fit Test}\label{psi}
\subsection{Problem Statement}
As described above, our randomization engine employs MD5 hashing and MOD functions \cite{kohavi2009controlled} (in equation~\ref{eq_hash}). It is important to check whether the distribution among traffic buckets is uniform and evenly distributed for a running experiment. 
We call it an "in-flight randomization validation" as it can be performed while an experiment is in progress. 

Given an experiment, we denote the sample size $n_b$ in bucket $b$ and total sample size $n=\sum_{b=0}^{B-1} n_b$, the goal is to test whether the sample distribution $\hat{p}_b=\frac{n_b}{n}$ matches the expected distribution $\frac{1}{B}$ in Multinomial when evenly distribution is by design. The hypotheses of the distribution test is
\begin{equation}
\label{eq_hypothesis}
\begin{aligned}
  \mathbf{H}_0: \forall b \in [0, B), p_b = \frac{1}{B} \\
  \mathbf{H}_a: \exists b \in [0, B), p_b \neq \frac{1}{B}
\end{aligned}
\end{equation}

To validate the randomization results while collecting samples, we can utilize industry standard statistical goodness-of-fit tests \cite{kohavi2020trustworthy,linkedin_chen2019b} as baselines, e.g., Pearson ${\chi}^2$ test \cite{chisq_cochran1952chi2,simard2011computing}, KS test \cite{massey1951kolmogorov} and AD test \cite{scholz1987_ad_test,razali2011power}.
The drawback of the widely adopted methods is that they generate non-ignorable false positives and false negatives that impact the effectiveness of the randomization validation process, i.e., either missing signals, or generating many false alerts.

Considering that our platform can run hundreds and thousands of experiments concurrently, an algorithm with a $0.5\%$ false positive rate may slow down the test pace for unnecessary manual investigation, resulting in delay and loss of product development opportunity. In addition to the standard algorithms, we developed a new method based on Population Stability Index (PSI \cite{psi_yurdakul2019statistical}).

\subsection{$\mathbf{PSI}$ and $\mathbf{PSI}_{k}$ Tests}

Population Stability Index (a.k.a. PSI) is computed as 
\begin{equation}
\label{eq_psi_t}
\begin{aligned}
  \mathbf{PSI} = \sum^{B-1}_{b=0} (\hat{p}_b - \hat{q}_b)\ln{\frac{\hat{p}_b}{\hat{q}_b}}
\end{aligned}
\end{equation}
where $\hat{p}_b$ and $\hat{q}_b$ are the sample proportions in the bucket $b$ (with total $B$ buckets) from two distributions $\mathbf{Multinom}(\mathbf{p})$ and $\mathbf{Multinom}(\mathbf{q})$. 

PSI is a test statistic \cite{psi_yurdakul2019statistical} that has been used to measure how much the distribution of a random variable has shifted over time, or to measure the distribution difference. Therefore, the original PSI is a two-sample hypothesis testing with $m$ as the other total sample size from the reference distribution $\mathbf{Multinom}(\mathbf{q})$ to be compared. 
In theorem, $\frac{1}{\frac{1}{n} + \frac{1}{m}} \mathbf{PSI}$ approximately follows a $\chi^2$ distribution with $B-1$ degrees of freedom\cite{psi_yurdakul2019statistical},
\begin{equation}
\label{eq_psi}
\begin{aligned}
 \mathbf{PSI} \sim (\frac{1}{n} + \frac{1}{m}) * \chi^{2}_{B-1}
\end{aligned}
\end{equation}
where $n$ and $m$ are the total sample size from two distributions $\mathbf{Multinom}(\mathbf{p})$ and $\mathbf{Multinom}(\mathbf{q})$ respectively. 
To distinguish the two total sample size, we use $m$ as the reference sample size and $n$ as the assigned sample size.

In the randomization validation case, we adapt PSI to determine if the assigned sample distribution over buckets $(n_0, \dots, n_{B-1})$ is the same as sampling from an uniform Multinomial distribution $(\frac{kn}{B}, \dots, \frac{kn}{B})$ with the reference total sample size $m=kn$. Here $k \in \mathbb{N}^{+}$ is the ratio of the reference sample size to the assigned total sample size. We found that $k$ could be regarded as a hyper parameter, and our algorithm performed slight better by setting $k=2$ as a practice trick.


From Equation \ref{eq_psi}, we can test whether the assigned sample distribution $\hat{p}_b=\frac{n_b}{n}$ are evenly distributed across $B$ buckets ($\hat{q}_b=\frac{1}{B}$) with a test statistic $\mathbf{PSI}_{k}$ defined as,
\begin{equation}
\label{eq_psi_k}
\begin{aligned}
  \mathbf{PSI}_{k} &= \sum^{B-1}_{b=0} (\frac{n_b}{n} - \frac{1}{B})(\ln{\frac{n_b}{n}} - \ln{\frac{1}{B}}) \\
  & \sim \frac{k+1}{kn} \chi^{2}_{B-1}
\end{aligned}
\end{equation}
where $k \in \mathbb{N}^{+}$ is a hyper parameter and can be interpreted as the ratio of the reference sample size to the assigned sample size $n$.

We can construct a decision rule along with significance level $\alpha$ for sending imperfect randomization alert when
\begin{equation}
\label{eq_psi_k_test}
\begin{aligned}
  \mathbf{PSI}_{k} & > \frac{k+1}{kn} \chi^{2}_{\alpha,B-1}
\end{aligned}
\end{equation}
where $\chi^{2}_{\alpha,B-1}$ is the critical value that has $(1-\alpha)\times 100\%$ area from the left tail of $\chi^{2}_{B-1}$ distribution with $B-1$ number of degrees of freedom.
As a trival example, when $k=1$, the decision region will be $\mathbf{PSI}_{k=1} > \frac{2}{n} \chi^{2}_{\alpha,B-1}$. 

Our precision and recall analysis and comparison in the following demonstrates that $\mathbf{PSI}_{k}$ performs the best both in our real-data and simulations. 
It is robust and can dramatically reduce false positive alarms in our production pipeline, which motivates us to share our findings with broader audience. 

\subsection{Performance Analysis}
The non-ignorable false positive rate is the fundamental concern with standard goodness-of-fit testings. We show that $\mathbf{PSI}$ can further suppress false positives when compared to KS test, the best randomization validation test so far in industry. 
We constructed three datasets from the real production data and simulation to prove our point, as summarized in Table \ref{tab0}. To evaluate the performance, we reported the false positive rate (FPR), precision, recall and F-score on three datasets.

\begin{table}[H]
  \begin{tabular}{|l|l|}
  \hline
  \multicolumn{1}{|c|}{Dataset}     & Description                                                                                   \\ \hline
  \multicolumn{1}{|c|}{1}   & 306 negative cases collected in real data \\ \hline
  \multicolumn{1}{|c|}{2}   & 291 negative cases collected in real data \\ 
                            & 33 positive cases with on average 0.09\% noise \\ \hline
  \multicolumn{1}{|c|}{3}   & 500 negative (evenly distributed) \\
                            & 100 positive cases with on average 0.09\% noise \\
                            & simulated based on Multinomial distribution                                                           \\ \hline
  \end{tabular}
  \caption{Datasets for Randomization Validation Comparison} \label{tab0}
  \vspace{-1.5em}
\end{table}

The Dataset 1 was collected from real online experiment metadata which contains no sample distribution anomaly, and all cases are considered negative. 

The Dataset 2 was based on the experiment metadata with no anomaly but randomly mixed with sample count anomalies, causing the buckets to be imbalanced. 
Specifically, we artificially added an extra $(0.05 + x10^{-2})\%$ of sample counts into certain bucket, where $x \sim Poisson(4)$. By doing so, we injected on average $0.09\%$ noise into certain buckets as the imbalanced cases which is aligned with our daily practice. 
This resulted in small but positive sample count distribution anomaly due to extra samples in buckets. 
In a summary, we generated 33 positive cases and kept 291 negative cases. 
For Dataset 1 and 2, the total sample size of collected experiments ranged from half million to several billion samples, which includes both small and big historical experiments in eBay.

The Dataset 3 is coming from simulation with negative and positive cases based on Multinomial distributions purely for reproduction purpose. 
Specifically, the sample count of a bucket is drawn from a Multinomial distribution with $p_b=\frac{1}{100}$, and total sample size $n \sim Poisson(3 \times 10^6)$. Similar to Dataset 2, we randomly picked up to 5 buckets to inject anomalies with new sample proportions in these buckets being $p_b=\frac{1}{B} + (0.05 + x10^{-2})\%$ where $x \sim Poisson(4)$, which injects on average $0.09\%$ noise.  
In total, we simulated 500 negative and 100 positive cases. 


We found that, in Table \ref{tab1}, AD test performed better than Pearson chi-square test, while KS test was the best out of the three. But there was still $0.33\%$ FPR for KS test. It translates to 3 alerts among 1000 concurrent experiments at any given time, which is barely acceptable. $\mathbf{PSI}_{k=1}$ had zero FPR in Dataset 1. This might be a coincidence, but it also suggested $\mathbf{PSI}_{k=1}$ could further silence false alerts comparing with KS test.
\begin{table}[H]
  \begin{tabular}{@{}|c|c c|c c|c c|c c|@{}}
  \hline
  \multirow{2}{*}{Dataset 1}    & \multicolumn{2}{c|}{${\chi}^2$ test} & \multicolumn{2}{c|}{AD test}  & \multicolumn{2}{c|}{KS test}  & \multicolumn{2}{c|}{$\mathbf{PSI}_{1}$ test} \\ \cline{2-9} 
                                & 0               & 1                  & 0               & 1           & 0               & 1           & 0              & 1                           \\ \hline
  True Label 0                  & 275             & 31                 & 302             & 4           & 305             & 1           & 306            & 0                           \\ \hline
  False Positive Rate           & \multicolumn{2}{c|}{$10.13\%$}       & \multicolumn{2}{c|}{$1.31\%$} & \multicolumn{2}{c|}{$0.33\%$} & \multicolumn{2}{c|}{$0\%$}                   \\ \hline
  \end{tabular}
  \caption{$\mathbf{PSI}$ performance evaluation on Dataset 1.} \label{tab1}
  \vspace{-1.5em}
\end{table}

Table \ref{tab2} also proved that the $\mathbf{PSI}_{k=1}$ outperformed all baselines with the highest F-score $98.46\%$. 
This is a significant achievement in practice, since our platform is expect to run thousands of experiments concurrently, even an $1\%$ improvement in reducing FPR can help us avoid unnecessary investigations. 
\begin{table}[H]
  \begin{tabular}{@{}|c|c c|c c|c c|c c|@{}}
  \hline
  \multirow{2}{*}{Dataset 2}    & \multicolumn{2}{c|}{${\chi}^2$ test} & \multicolumn{2}{c|}{AD test}  & \multicolumn{2}{c|}{KS test}  & \multicolumn{2}{c|}{$\mathbf{PSI}_{1}$ test} \\ \cline{2-9} 
                                & 0               & 1                  & 0               & 1           & 0               & 1           & 0              & 1                             \\ \hline
  True Label 0                  & 261             & 30                 & 287             & 4           & 290             & 1           & 291            & 0                             \\ 
  True Label 1                  & 0               & 33                 & 3               & 30          & 3               & 30          & 1              & 32                            \\ \hline
  FPR                 (in $\%$) & \multicolumn{2}{c|}{$10.31\%$}       & \multicolumn{2}{c|}{$1.37\%$} & \multicolumn{2}{c|}{$0.34\%$} & \multicolumn{2}{c|}{$0\%$}                     \\ \hline
  Precision           (in $\%$) & \multicolumn{2}{c|}{$52.38\%$}       & \multicolumn{2}{c|}{$88.24\%$}& \multicolumn{2}{c|}{$96.77\%$}& \multicolumn{2}{c|}{$100\%$}                   \\ \hline
  Recall              (in $\%$) & \multicolumn{2}{c|}{$100\%$}         & \multicolumn{2}{c|}{$90.11\%$}& \multicolumn{2}{c|}{$90.91\%$}& \multicolumn{2}{c|}{$96.97\%$}                 \\ \hline
  F-score             (in $\%$) & \multicolumn{2}{c|}{$68.75\%$}       & \multicolumn{2}{c|}{$89.55\%$}& \multicolumn{2}{c|}{$93.75\%$}& \multicolumn{2}{c|}{$98.46\%$}                 \\ \hline
  \end{tabular}
  \caption{$\mathbf{PSI}$ performance evaluation on Dataset 2.} \label{tab2}
  \vspace{-1.5em}
\end{table}

Table \ref{tab3} showed that $\mathbf{PSI}$ performed similarly in simulated Dataset 3.
$\mathbf{PSI}_{k=1}$ testing was found to be sensitive and accurate in both precision and recall measures. Readers are welcome to reproduce the results by themselves.
\begin{table}[H]
  \begin{tabular}{@{}|c|c c|c c|c c|c c|@{}}
  \hline
  \multirow{2}{*}{Dataset 3}    & \multicolumn{2}{c|}{${\chi}^2$ test} & \multicolumn{2}{c|}{AD test}  & \multicolumn{2}{c|}{KS test}  & \multicolumn{2}{c|}{$\mathbf{PSI}_{1}$ test} \\ \cline{2-9} 
                                & 0               & 1                  & 0               & 1           & 0               & 1           & 0              & 1                             \\ \hline
  True Label 0                  & 447             & 53                 & 494             & 6           & 496             & 4           & 500            & 0                             \\ 
  True Label 1                  & 0               & 100                & 22              & 78          & 27              & 73          & 0              & 100                           \\ \hline
  FPR                 (in $\%$) & \multicolumn{2}{c|}{$10.60\%$}       & \multicolumn{2}{c|}{$1.20\%$} & \multicolumn{2}{c|}{$0.80\%$} & \multicolumn{2}{c|}{$0\%$}                     \\ \hline
  Precision           (in $\%$) & \multicolumn{2}{c|}{$65.36\%$}       & \multicolumn{2}{c|}{$92.86\%$}& \multicolumn{2}{c|}{$94.81\%$}& \multicolumn{2}{c|}{$100\%$}                   \\ \hline
  Recall              (in $\%$) & \multicolumn{2}{c|}{$100\%$}         & \multicolumn{2}{c|}{$78.00\%$}& \multicolumn{2}{c|}{$73.00\%$}& \multicolumn{2}{c|}{$100\%$}                   \\ \hline
  F-score             (in $\%$) & \multicolumn{2}{c|}{$79.05\%$}       & \multicolumn{2}{c|}{$84.78\%$}& \multicolumn{2}{c|}{$82.49\%$}& \multicolumn{2}{c|}{$100\%$}                   \\ \hline
  \end{tabular}
  \caption{$\mathbf{PSI}$ performance evaluation on Dataset 3.} \label{tab3}
  \vspace{-1.5em}
\end{table}
The noise scale $0.09\%$ is discussed according to practice, while our proposed algorithm actually perform very well under different noise scale. 
In Appendix~\ref{psi_data_appendix}, we report the performance under different noise scale from $0.05\%$ to $0.15\%$ on average. 

As a trick in practice, we could fine-tune the performance of $\mathbf{PSI}_{k}$ by varying with different $k$ values for different use cases.
Table \ref{tab4} demonstrated an example of $k$ tuning using grid search over $k$ and summarized the results in our datasets. 
We discovered that, by increasing $k$, we may have higher recall while could damage precision. 
We finally settled with $k=2$ in eBay according to the human diagnose feedback of detected experiments.
While, it is still an interesting open question on tuning the $k$ from the data.


\begin{table*}[!ht]
  \centering
  \begin{tabular}{@{}|l|c|c|c|c|c|c|c|c|@{}}
  \hline
  \multicolumn{1}{|c|}{Dataset} & \multicolumn{1}{|c|}{$\mathbf{PSI}_{k}$ test}   & $k=1$       & $k=2$       & $k=3$        & $k=4$        & $k=5$       & $k=6$       & $k=7$        \\ \hline
  \multirow{1}{*}{Dataset 1}    & \multicolumn{1}{|c|}{False Positive Rate}       & $0.00\%$    & $0.00\%$    & $0.00\%$     & $0.00\%$     & $0.33\%$    & $0.33\%$    & $0.98\%$     \\ \hline
  \multirow{3}{*}{Dataset 2}    & \multicolumn{1}{|c|}{False Positive Rate}       & $0.00\%$    & $0.00\%$    & $0.00\%$     & $0.00\%$     & $0.34\%$    & $0.34\%$    & $1.03\%$     \\ \cline{2-9}
                                & \multicolumn{1}{|c|}{Precision}                 & $100\%$     & $100\%$     & $100\%$      & $100\%$      & $97.06\%$   & $97.06\%$   & $91.67\%$    \\ \cline{2-9}
                                & \multicolumn{1}{|c|}{Recall}                    & $96.97\%$   & $100\%$     & $100\%$      & $100\%$      & $100\%$     & $100\%$     & $100\%$      \\ \hline
  \multirow{3}{*}{Dataset 3}    & \multicolumn{1}{|c|}{False Positive Rate}       & $0.00\%$    & $0.00\%$    & $0.00\%$     & $0.00\%$     & $0.00\%$    & $0.20\%$    & $0.60\%$     \\ \cline{2-9}
                                & \multicolumn{1}{|c|}{Precision}                 & $100\%$     & $100\%$     & $100\%$      & $100\%$      & $100\%$     & $99.01\%$   & $97.09\%$    \\ \cline{2-9}
                                & \multicolumn{1}{|c|}{Recall}                    & $100\%$     & $100\%$     & $100\%$      & $100\%$      & $100\%$     & $100\%$     & $100\%$      \\ \hline
  \end{tabular}
  \caption{$\mathbf{PSI}_{k}$ performance evaluation varying $k=1, 2, 3, 4, 5, 6,7 $} \label{tab4}
  \vspace{-1.0em}
\end{table*}
In a summary, $\mathbf{PSI}_{k}$ is a viable solution along with other well-known methods with on-par or better performance on precision and recall. It will be interesting for other researchers to confirm its performance on other data sets. $\mathbf{PSI}_{k}$ is an indispensible tool to improve the effectiveness of our randomization engine. 

\subsection{Randomization Case Study}
We have implemented $\mathbf{PSI}_{k}$ in our experimentation platform for the randomization "in-flight" validation monitoring. In practice, we occasionally observe imperfect randomization results (samples are not evenly distributed among buckets) due to "unhappy randomization" and operation issues. A operation issue on experience tracking and randomization engine is severe as it could corrupt more than one experiment. After our deployment of $\mathbf{PSI}_{k}$, we can catch the imperfect randomization issues at a very early stage to take actions, instead of dealing with experimenters' complains at the end of data collection. To be honest, issue from the randomization engine is usually rare. Below, we share two cases here, in one case it helped us catch a bug.

\paragraph{Case 1. Sample Leakage Detection} \

\textbf{Experiment}: In 2021, one domain team ran an experiment with 10\%-10\% traffic split. The samples with mod values 0-9 were assigned to the control and 10-19 to the test group in the experiment setting, while the rest samples with mod values 20-99 are unassigned traffic buckets. Shortly after our $\mathbf{PSI}_{k}$ deployment, it caught an imbalanced sample count among the 0-99 buckets in this experiment. Moreover, there exists more samples in each bucket with mod values 0-19 than 20-99, where some "ghost" users accidentally leaked into two or more buckets. In the triggered sample report, we observed about $0.3\%$ of "ghost" users with mod values 20-99 appeared in either test or control group. Interestingly, these "ghost" users were uniformly dispatched among the unassigned traffic buckets.

\textbf{Explanation}: After further research, we concluded that the incident was not caused by a flaw in our analytic data pipeline, but rather by a malfunction on the experience service side. A tracking application recorded incorrect user ids (cookies) at a specific point in the service pool. The cookie ids recorded in the logs were different from those fed into the randomization engine. This could happen as the application wrongly decide which user ids (cookies) to track the user given the situation at a specific point and time during the experiment period. It appeared as if some users accidentally leaked into different randomization buckets in an experiment, and caused a randomization anomaly. $\mathbf{PSI}_{k}$ monitoring was able to successfully detect the sample imbalance early to initiate an investigation. It is important to note that if one of the buckets belongs to the control and another belongs to the test group, it may trigger sample delta alert on the triggered samples which we discuss in the next section. While such mistakes could pollute several experiments together, it is important to alarm earlier on total traffic assignment instead of waiting until multiple experiments triggered sample delta alert.

\paragraph{Case 2. Unhappy Randomization} \

Because the hashing and MOD functions aren't perfect, the randomization output will not always meet an experiment's requirement. Researchers have discovered that choosing the proper randomization seed reduces the risk of sample size differences. Re-randomization is a technique for reducing the noise produced by "unhappy randomization" and improving the test precision \cite{morgan2012rerandomization}. As $\mathbf{PSI}_{k}$ can identify unhappy randomization early, we may warn experimenters to re-randomize samples when seeing imbalanced buckets using a different randomization seed.

\section{Sample Ratio Mis-Match Monitoring}\label{sprt}
\subsection{Delta among Triggered Sample Counts}
Imagine in an experiment, 50 buckets on a plane are assigned to the test group and 50 buckets to the control group. The test and control groups should have roughly the same number of samples. However we may still observe a noticeable or significant difference between the observed sample count from the expected value, e.g. there may be 1\% more samples in the test group than that of the control group. In our experience, a 1\% deviation is more surprising with large samples but may be expected on small samples. We call this “sample delta” or sample ratio mismatch in the literature. 

Although a small difference is always expected due to tracking data time delay and traffic fluctuation, the magnitude of sample delta issues can raise concerns. We developed a sample delta detection method based on sequential analysis with a significance threshold to monitor all live experiments. It becomes one of the most critical health checks in our experimentation platform.  For both the assigned traffic and triggered traffic of each experiment, we calculate sample delta respectively for diagnostic and validation purposes.
As the method can be applied in both assigned (randomization only) and triggered sample sets, we use the triggered samples to illustrate the principle of the method.
\subsection{Sample Delta Sequential Test}
Let $x_1^{T}, \dots, x_{k}^{T}$ and $x_1^{C}, \dots, x_{k}^{C}$ be the cumulative sample count in the control (C) and treatment (T) groups of an experiment at day $k$ with assigned traffic $r^{T}$ and $r^{C}$. We can construct a two-sided hypothesis test under binomial setting with observed sample ratio $p = \frac{x_{k}^{T}}{x_{k}^{T} + x_{k}^{C}} $ 
and expected sample ratio $p_0 = \frac{r^{T}}{r^{T} + r^{C}}$: 
\begin{align}
H_{0}: p - p_0& = 0  \notag \\
H_{A}: p - p_0 & \neq 0  \notag
\end{align}

The naive design hinges on a proportion metric illustrated in the above hypothesis. As proportional t-tests are usually very sensitive, the hope is to catch the sample delta as soon as possible. In fact, it turns out the t-test is too sensitive especially when the sample size $n_{k}=x_{k}^{T} + x_{k}^{C}$ is large, as it will almost certainly detect a small but negligible difference (e.g. $0.1 \%$) throughout the days of an experiment with millions of samples. But a $0.1 \%$ deviation with millions samples is very common in our daily practice and should not raise concern on our system. In the meanwhile, we still expect to get alert for big sample delta (e.g. $1\%$) with moderate sample size.

So we change our sample delta detection test hypothesis to include a threshold as following:
\begin{align}
H_{0}:  \lvert p - p_0 \rvert\ & \leq  \delta  \notag \\
H_{A}: \lvert p - p_0 \rvert\ & \geq \delta  \notag
\end{align}
where $\delta$ is an error tolerance hyper parameter that can be tuned to surpress the noises. To start with, we set $\delta = \min(1\%, 5\% min(p_0, 1 - p_0))$.

Since we are going to run the same sample delta test repeatedly for each experiment at an hourly or daily basis, it would definitely inflate the false positive (type-I error) rate for the detection. Instead of conducting a proportional t-test, we design the detection based on the sequential probability ratio test (SPRT) \cite{waldseq} to minimize the inflated type-I errors. 
Specifically, we design two adaptive SPRT tests (one-sided):
\begin{align*}
&Test A &Test B \\
&H_{0}:  p - p_0  =  0   & H_{0}:   p_0 - p =  0  \\
&H_{A}: p - p_0  \geq \delta  & H_{A}:   p_0 - p  \geq \delta
\end{align*}
with the corresponding test statistics (see Appendix \ref{proof1}) as:
\begin{align}
  t^{A}_{k} = \log \frac{l_{H_{A}}}{l_{H_{0}}}= - \frac{\delta^2 - 2 \delta(\hat{p}_{k} - p_0) }{2\hat{\sigma}_{k}^2}  \label{t1} \\
  t^{B}_{k} = \log \frac{l_{H_{A}}}{l_{H_{0}}}= -\frac{\delta^2 + 2 \delta(\hat{p}_{k} - p_0) }{2\hat{\sigma}_{k}^2} \label{t2}
\end{align}

where $\hat{p}_{k} = \frac{x_{k}^{T}}{x_{k}^{T} + x_{k}^{C}}$ (the symbol $\hat{*}$ means it is computed in a real example) and $\hat{\sigma}_{k}^2 = \frac{\hat{p}_{k} (1 - \hat{p}_{k} )}{n_{k}}$. A more accurate version of the formula (SPRT-EXACT) using binomial distribution directly instead of the normal distribution via CLT (see Appendix \ref{proof2}) can be:

\begin{align}
  t^{A}_{k} =  x^{T}_{k}log(\frac{p_0 + \delta}{p_0}) + x^{C}_{k}log(\frac{1 - p_0 - \delta}{1 - p_0}) \label{t3} \\
  t^{B}_{k} = x^{T}_{k}log(\frac{p_0 - \delta}{p_0}) + x^{C}_{k}log(\frac{1 - p_0 + \delta}{1 - p_0}) \label{t4}
\end{align}

The alerting rules of the SPRT tests are based on the critical region \citep{waldseq} with type-I $\alpha$ and type-II $\beta$ error bounds.
For any $k =1,\dots,\infty$:
\begin{itemize}
\item if $t_{K} > log(\frac{1 - \beta}{\alpha})$:  \\
 Send alert and break;
\item if $t_{K} < log(\frac{1 - \beta}{\alpha})$: \\
 Continue monitoring;
 \end{itemize} 
 
\subsection{Performance Analysis}
We choose the t-test and chi-square tests as the methodological comparison baseline as they are the most commonly used, to illustrate the performance of the SPRT. The validation dataset is based on 519 long running experiments with average K = 29 days data points. 

The dataset is labeled with the criterion that  if $\frac{|x_k^T - x_k^C|}{x_k^C}$ $>$ $1\%$ as the evidence of sample delta existed. The label is reflected in the following table \ref{tab5}, where 0 indicates that there was no sample delta issue and 1 indicates that there was a sample delta issue. According to our experience, the rule may result in small amount of false positives, particularly  when sample size is small, but no false negative.

\begin{figure}[!hb]
  \includegraphics[width=0.9\linewidth]{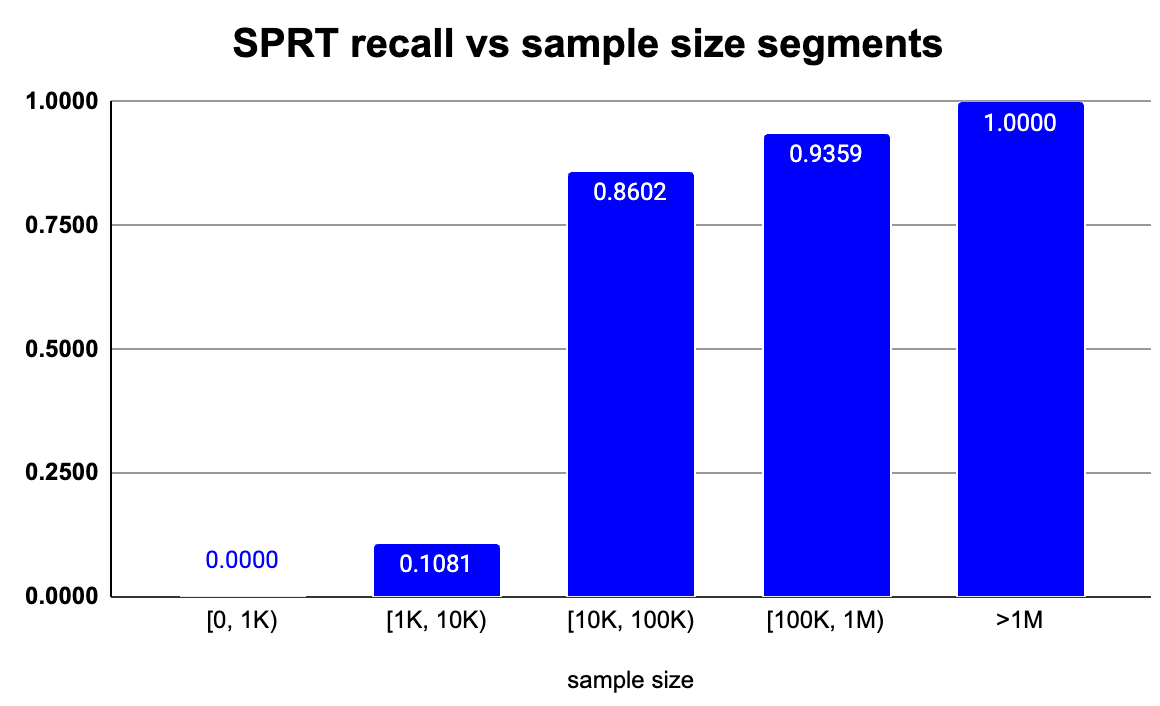}
  \caption{ SPRT recall rate in different sample size segments. }
  \label{fig:recall}
  \vspace{-1.0em}
\end{figure}
\clearpage

We report the precision as the primary evaluation metric as well as recall in our confusion table (Table \ref{tab5}). We choose $\alpha = 0.05$, $\beta=0$ for SPRT and SPRT-EXACT methods. We also intentionally lower $\alpha$ to 0.01 for the chi-square and t-test statistics to further reduce their family wise errors.
\begin{table}[!ht]
  \centering
  \begin{tabular}{@{}|c|c c|c c|c c|c c|@{}}
  \hline
  \multirow{2}{*}{}    & \multicolumn{2}{c|}{\begin{tabular}[c]{@{}c@{}} ${\chi}^2$ test\\ ($\alpha = 1\%$) \end{tabular}} & \multicolumn{2}{c|}{\begin{tabular}[c]{@{}c@{}} t-test\\ ($\alpha = 1\%$) \end{tabular}}  & \multicolumn{2}{c|}{\begin{tabular}[c]{@{}c@{}} \textbf{SPRT}\\ ($\alpha =5\%$) \end{tabular}}  & \multicolumn{2}{c|}{\begin{tabular}[c]{@{}c@{}} \textbf{SPRT-EXACT}\\ ($\alpha = 5\%$) \end{tabular}} \\ \cline{2-9} 
                                & 0               & 1                  & 0               & 1           & 0               & 1           & 0              & 1                             \\ \hline
  True Label  0                  & 13798             & 719                 & 13798             & 719           & 14501             & 16           & 14501            & 16                             \\ 
  True Label  1                  & 309               & 285                & 307              & 283          & 310              & 282          & 310              & 282                           \\ \hline  Precision           & \multicolumn{2}{c|}{$28.39\%$}       & \multicolumn{2}{c|}{$28.24\%$}& \multicolumn{2}{c|}{$\mathbf{94.63\%}$}& \multicolumn{2}{c|}{$\mathbf{94.63\%}$}                   \\ \hline
  Recall              & \multicolumn{2}{c|}{$47.98\%$}         & \multicolumn{2}{c|}{$47.97\%$}& \multicolumn{2}{c|}{$47.64\%$}& \multicolumn{2}{c|}{$47.64\%$}                   \\ \hline
  \end{tabular}
  \caption{Sample delta detection method comparison} \label{tab5}
  \vspace{-1.5em}
\end{table}

Table \ref{tab5} indicates that both SPRT and SPRT-EXACT can dramatically increase alert precision from $~28\%$ to $~95\%$. Actually SPRT and SPRT-EXACT have identical results. The SPRT and SPRT-EXACT tests appear to be superior to the direct t and chi-square tests.

It's worthy to note that the rule-based labels are used in all four methods, resulting in a $~48\%$ recall rate.   We plot a bar graph of the recall rate versus sample sizes in Fig. \ref{fig:recall}. The graph illustrates that as the sample size grows, the SPRT greatly boosts the recall rate. This could be partially explained by rule-based labeling' significant false positives in small samples. We maintain that the key focus here is to the significant reduction of the type-I error when monitoring a large number of online A/B tests in an experimentation platform.

\subsection{Sample Delta Debugging and Case Study}

Our experimentation platform implements the automated detection and monitoring process for sample delta troubleshooting. The proposed Sample Delta method has been monitoring the triggered samples and these samples under different user cohorts for each eBay experiment since early 2019. It provides insights by the following aspects whenever a sample delta is detected for an experiment:

 \begin{enumerate}
  \item By date: did the sample delta appear only after a certain date? 
  \item By the triggered sample set or assigned sample set: did the sample delta appear only in the triggered traffic or in the MOD assigned and qualified traffic as well? 
  \item By the impact on other experiments: did any other experiments suffer sample delta issue on/from the same day as well? Did any experiment have any significant metric change starting from the start day of the sample delta issue? 
  \item By site/channel/browser families: did the sample delta only appear on certain sites, channels(Android, iOS, Web etc.), certain types of browsers?
\end{enumerate}

Information from different aspects can be combined to generate deeper insight to help troubleshoot the sample delta issue. As shown in Figure \ref{fig:srm}, the sample delta detection evaluates the cumulative sample counts at both assigned and triggered level for each live experiment by all the aspects mentioned above to see if a sample delta is significant. If the sample delta is significant for an experiment, the experiment owner must identify what the root cause is, fix the error and re-run the experiment. The faster the platform notices the problem, the earlier the experimenter can take action. Most of the sample ratio mismatch cases caused by underlying issues have large differences that don't require a significant amount of samples to detect. As a result, a real-time data pipeline has also been implemented in the platform in order to detect these issues more quickly.

The procedure of troubleshooting a sample delta issue is arduous because there can be many probable causes. Most of the root causes lie outside of the experimentation platform. The platform can assist by narrowing down the problem space to facilitate the debugging process.  For example, if the sample delta appears only in the triggered traffic after a particular date but not in the assigned traffic, it may indicate a code release on the experimenter's development side, with an implementation issue propagated to the triggered traffic on that date. In the following, we review two real-life examples that we came across in practice.


\paragraph{Case 1. Bot Filtering of Online Traffic}  \

\textbf{Experiment}: In 2021, one domain team observed sample delta alerts at both the triggered and assigned traffics, shortly after the beginning of the experiment. The sample delta was much larger in the triggered traffic ($\approx 5\%$) than that from the assigned traffic ($\approx 3\%$). The sample delta was mostly flat (~$\approx 0\%$) in many country sites except significant  ($\approx 11\%$) in a particular country \textbf{F}. In the meanwhile, for many other experiments running in the particular country \textbf{F}, the total samples and the user sessions started to increase right after the problematic experiment started to run. The experimentation platform detected the sample delta ratio mismatch and alerted the experiment owners. 

\begin{figure}[!hb]
  \includegraphics[width=0.9\linewidth]{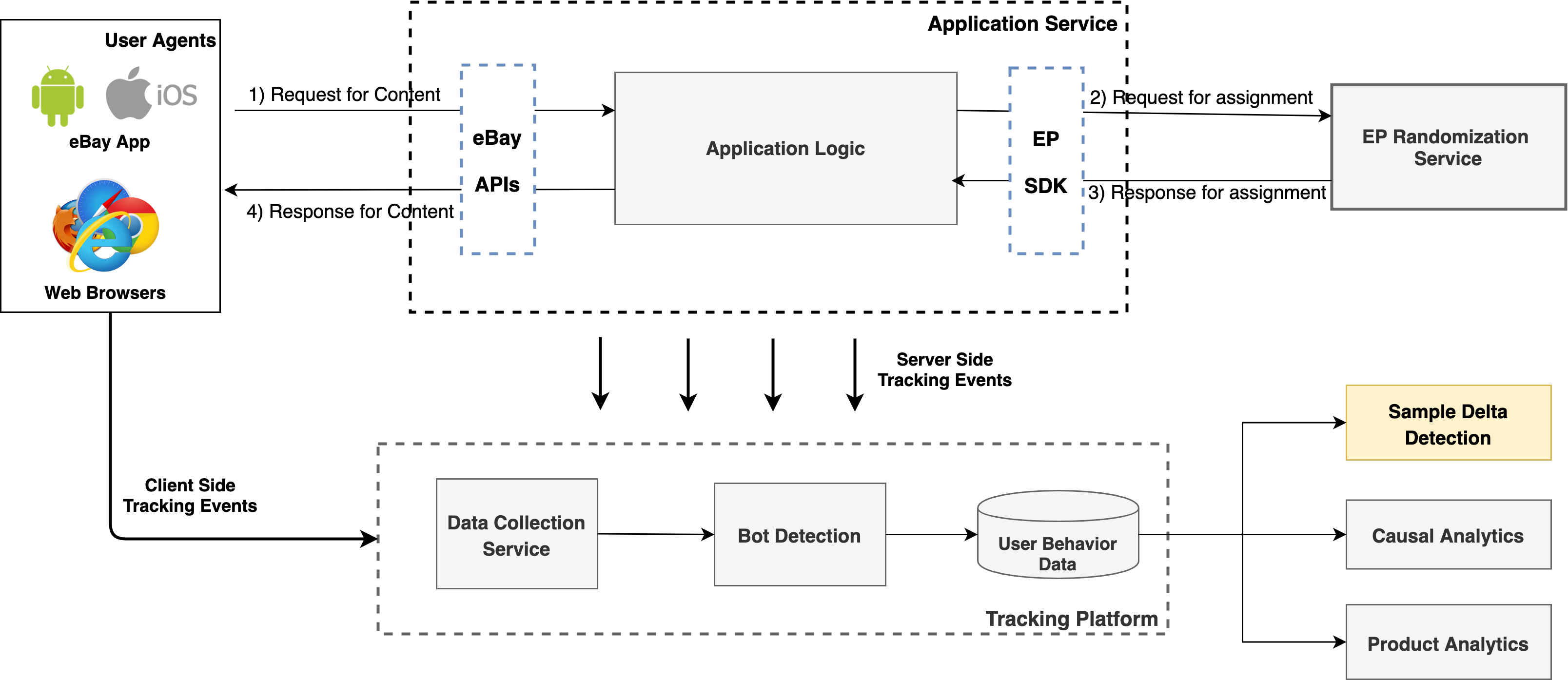}
  \caption{Experiment and Tracking Data Flow}
  \label{fig:data_flow}
  \vspace{-1.0em}
\end{figure}

\textbf{Explanation}: After a comprehensive diagnose, we conclude that the test variant caused a latency issue on the particular country \textbf{F}'s site. This latency messed up with system bolt filtering logic and hence broke the traffic balance. Online application tracking events usually cannot be consumed directly by analytics and experiments without going through proper bot detection. Figure \ref{fig:data_flow} illustrates tracking data with experimentation tags can be generated from the customer's client side and the online server side applications. It is then written into Kafka pools. The tracking data should be processed such as to filter out the external bots and the internal server side events, only genuine customer traffic are allowed to flow through to the downstream applications.

\clearpage

At the time, one of the common rules of bot detection would recognize a visitor as a bot if it had too many user activities in a short period of time. In this case, when the test variant slowed down the traffic, a significant amount of the bot traffic would not be recognized as bots because the activities took longer time. As such the experiment saw the imbalanced traffic between test and control due to additional un-filtered bot traffic in the test group. For other experiments, the session count and sample count also increased after the experiment started.

\paragraph{Case 2. Online Experience Redirect} \

\textbf{Experiment}: In 2021, a development team rewrote their service on a new tech stack (technology migrations). Then, they ran an experiment to see the impact on metrics to compare the old with the new service. The traffic hit the old service directly as it was in production. However for a user in the test group, the experiment redirected the traffic to the new service. A sample delta at the triggered sample level was observed immediately after the experiment started but the randomization assigned traffic was flat. 

\textbf{Explanation}: It turned out the additional hop of the redirection was the culprit. Additional breakdown on service identifier showed that the assigned traffic on the old service is flat, but the amount of assigned test traffic on the new service is much less than the assigned traffic on the old service in control. The redirection produced a non-negligible failure rate and delays, and resulted in a sample ratio mismatch alert in triggered traffic. The team re-implemeted the logic and restarted the experiment without a further issue. 

\section{Summary and Future Work} \label{smry}
In the paper, we describe two methods and processes to ensure A/B test quality at the sample assignment and triggering stages for experimentation needs. The novel population stability index (PSI) test method and the sequential analysis algorithm enable the ability to automatically monitor potential randomization issues and sample discrepancies among the control and test groups for all experiments at scale with the required sensitivity but minimized false alerts. We deployed the two methods in experimentation platform for monitoring and alerting the health of every experiment in eBay. They are critical for the platform to successfully execute thousands of high quality and trustworthy experiments to promote the experimentation culture among product developers. In the future, we plan to implement other methods to continuously monitor experiment pre-existing bias in decision metrics, e.g., accumulative and ratio metrics \citep{nie2020dealing}. Besides, testing immature features could cause outages or critical harmness on key business metrics. The monitoring on key metrics \citep{xu2018sqr,zhao2018safely,zhang2020moving} can help alarm experimenter at the very early hours and ensure a safe data collection, which can be another interesting application to further explore.

\bibliographystyle{ACM-Reference-Format}
\bibliography{reference}

\appendix
\section{Sequential Probability Ratio Test} \label{proof}
Consider a random variable $Y$ with density function $f(y|p)$ parametrized by $p$. For the simple hypothesis:
  \begin{align*}
&H_{0}:  p  =  p_0    \\
&H_{A}: p  \geq  p_1  
\end{align*}
based on $n$ independent observations $y_1, \dots, y_n$. The log likelihood ratio is defined as:
\begin{equation*}
\lambda_N = \log \frac{l_{H_{A}}}{l_{H_{0}}} = \log \frac{\prod_{j=1}^{n}f_1(y_j)}{\prod_{j=1}^{n}f_0(y_j)}
\end{equation*} 
where $l_{H_{\cdot}}$ stands for the likelihood of observed samples based on distribution in $H_{\cdot}$, To simplify the notation, let $f_1 = f(y| p_1 )$ and $f_0 = f(y| p_0 )$. According to Neyman-Pearson theory (\citep{lehmann2005testing}), the most powerful test is given as the following:
\begin{align*}
\textbf{Reject $H_0$ if } & \lambda_n > r \\
\textbf{Accept $H_0$ if } & \lambda_n \leq r \\
\end{align*}

The idea of sequential likelihood ratio testing is to add a third possibility in addition to accepting or rejecting, we can also elect to go on collecting data and decide later. Specifically, choose constants $A < B$ and sample $y_1, y_2, \dots$ sequentially until the random stopping time
\begin{equation*}
n_k = \min \{ k:\lambda_k \notin [A, B] \}
\end{equation*}
at which point we reject  $H_0$ if $\lambda_{n_k} > B$ and accept $H_0$ if $\lambda_{n_k} < A$. Wald \citep{waldseq} further showed that we can set $A = \log \frac{\beta}{1-\alpha}$ and $B = \log \frac{1-\beta}{\alpha}$ to control the overall type I error $\alpha$ and type II error $\beta$.

\subsection{Gaussian case} \label{proof1}
In the Gaussian case, we assume $\hat{p}_{k} = \frac{x^{T}_{k}}{n_{k}}$ follows distribution $N(p_0, \sigma^2)$ under $H_0$, and $N(p_0 + \delta, \sigma^2)$ under $H_A$.  The likelihood is given by
\begin{equation*}
\frac{1}{\sqrt{2\pi\sigma^2}} \exp{ \left[-\frac{1}{2\sigma^2} (\hat{p}_{k} - p)^2 \right]}
\end{equation*}
Then the log likelihood ratio is:
\begin{align*}
\lambda_k &= \log\frac{\frac{1}{\sqrt{2\pi\sigma^2}} \exp{ \left[-\frac{1}{2\sigma^2} (\hat{p}_{k} - p_0 - \delta)^2 \right] }}{\frac{1}{\sqrt{2\pi\sigma^2}} \exp{ \left[-\frac{1}{2\sigma^2} (\hat{p}_{k} - p_0)^2 \right] }} \\
& = -\frac{1}{2\sigma^2} (\hat{p}_{k} - p_0 - \delta)^2 + \frac{1}{2\sigma^2} (\hat{p}_{k} - p_0)^2 \\
& = - \frac{\delta^2 - 2 \delta(\hat{p}_{k} - p_0) }{2\sigma^2} 
\end{align*}
Plug in the estimator $\hat{\sigma}_{k}^2 = \frac{\hat{p}_{k} (1 - \hat{p}_{k} )}{n_{k}}$ for $\sigma^2$, we observe the formula in equation \ref{t1} : $t^{A}_{k} =  - \frac{\delta^2 - 2 \delta(\hat{p}_{k} - p_0) }{2\hat{\sigma}_{k}^2}$. Similarly we can also obtain $t^{B}_{k}$ in equation \ref{t2} as well.

\subsection{Binomial case} \label{proof2}
In the binomial case, we assume $x_k$ following binomial distribution $B(n_k, p)$. The likelihood will be:
\begin{equation*}
C_{x_k}^{n_k} p^{x_k}(1-p)^{n_k - x_k}
\end{equation*} Suppose $p_0 = p_0$ and $p_1 = p_0 + \delta$. Then the log likelihood ratio is :
\begin{align*}
\lambda_k &= \log \frac{ C_{x^{T}_k}^{n_k} (p_0 + \delta)^{x^{T}_k}(1-p_0-\delta)^{x^{C}_k} }{ C_{x^{T}_k}^{n_k} p_0^{x^{T}_k}(1-p_0)^{x^{C}_k} } \\
&= x^{T}_{k}\log(\frac{p_0 + \delta}{p_0}) + x^{C}_{k}\log(\frac{1 - p_0 - \delta}{1 - p_0})
\end{align*}
So we obtain the formula in equation \ref{t3}: $t^{A}_{k} =  x^{T}_{k}log(\frac{p_0 + \delta}{p_0}) + x^{C}_{k}log(\frac{1 - p_0 - \delta}{1 - p_0})$. Similarly we could obtain equation \ref{t4} as well.

\section{Simulation on different noise scale} 
\label{psi_data_appendix}

We use same simulation settings in the Dataset 3, except we sample $x \sim Poisson(\lambda)$ with different $\lambda \in \{0,1,2,3,4,5,6,7,8,9,10\}$. 
Hence, we can inject average noise from $0.05\%$ to $0.15\%$ to show the excellent performance of $\mathbf{PSI}_{k=1}$ test under different noise scale.

\begin{table}[H]
  \begin{tabular}{@{}|c|c c|c c|c c|c c|@{}}
  \hline
                & \multicolumn{2}{c|}{${\chi}^2$ test} & \multicolumn{2}{c|}{AD test}  & \multicolumn{2}{c|}{KS test}  & \multicolumn{2}{c|}{$\mathbf{PSI}_{1}$ test} \\ \cline{1-9} 
  0.05$\%$ noise & \multicolumn{2}{c|}{$100\%$}       & \multicolumn{2}{c|}{$23.0\%$} & \multicolumn{2}{c|}{$19.0\%$} & \multicolumn{2}{c|}{$73.0\%$}                     \\ \hline
  0.06$\%$ noise & \multicolumn{2}{c|}{$100\%$}       & \multicolumn{2}{c|}{$38.0\%$}& \multicolumn{2}{c|}{$31.0\%$}& \multicolumn{2}{c|}{$77.0\%$}                   \\ \hline
  0.07$\%$ noise & \multicolumn{2}{c|}{$100\%$}       & \multicolumn{2}{c|}{$44.0\%$}& \multicolumn{2}{c|}{$41.0\%$}& \multicolumn{2}{c|}{$93.0\%$}                   \\ \hline
  0.08$\%$ noise & \multicolumn{2}{c|}{$100\%$}       & \multicolumn{2}{c|}{$54.0\%$}& \multicolumn{2}{c|}{$50.0\%$}& \multicolumn{2}{c|}{$94.0\%$}                   \\ \hline
  0.09$\%$ noise & \multicolumn{2}{c|}{$100\%$}       & \multicolumn{2}{c|}{$78.0\%$}& \multicolumn{2}{c|}{$73.0\%$}& \multicolumn{2}{c|}{$100\%$}                   \\ \hline
  0.1$\%$ noise & \multicolumn{2}{c|}{$100\%$}       & \multicolumn{2}{c|}{$75.0\%$}& \multicolumn{2}{c|}{$75.0\%$}& \multicolumn{2}{c|}{$100\%$}                   \\ \hline
  0.11$\%$ noise & \multicolumn{2}{c|}{$100\%$}       & \multicolumn{2}{c|}{$75.0\%$}& \multicolumn{2}{c|}{$72.0\%$}& \multicolumn{2}{c|}{$100\%$}                   \\ \hline
  0.12$\%$ noise & \multicolumn{2}{c|}{$100\%$}       & \multicolumn{2}{c|}{$73.0\%$}& \multicolumn{2}{c|}{$75.0\%$}& \multicolumn{2}{c|}{$100\%$}                   \\ \hline
  0.13$\%$ noise & \multicolumn{2}{c|}{$100\%$}       & \multicolumn{2}{c|}{$82.0\%$}& \multicolumn{2}{c|}{$82.0\%$}& \multicolumn{2}{c|}{$100\%$}                   \\ \hline
  0.14$\%$ noise & \multicolumn{2}{c|}{$100\%$}       & \multicolumn{2}{c|}{$86.0\%$}& \multicolumn{2}{c|}{$89.0\%$}& \multicolumn{2}{c|}{$100\%$}                   \\ \hline
  0.15$\%$ noise & \multicolumn{2}{c|}{$100\%$}       & \multicolumn{2}{c|}{$93.0\%$}& \multicolumn{2}{c|}{$89.0\%$}& \multicolumn{2}{c|}{$100\%$}                   \\ \hline
  \end{tabular}
  \caption{Recall on different noise scale.}
  \vspace{-1.5em}
\end{table}

\begin{table}[H]
  \begin{tabular}{@{}|c|c c|c c|c c|c c|@{}}
  \hline
                & \multicolumn{2}{c|}{${\chi}^2$ test} & \multicolumn{2}{c|}{AD test}  & \multicolumn{2}{c|}{KS test}  & \multicolumn{2}{c|}{$\mathbf{PSI}_{1}$ test} \\ \cline{1-9} 
  0.05$\%$ noise & \multicolumn{2}{c|}{$63.69\%$}       & \multicolumn{2}{c|}{$82.14\%$} & \multicolumn{2}{c|}{$86.36\%$} & \multicolumn{2}{c|}{$100\%$}                     \\ \hline
  0.06$\%$ noise & \multicolumn{2}{c|}{$66.23\%$}       & \multicolumn{2}{c|}{$86.36\%$}& \multicolumn{2}{c|}{$92.87\%$}& \multicolumn{2}{c|}{$100\%$}                   \\ \hline
  0.07$\%$ noise & \multicolumn{2}{c|}{$68.02\%$}       & \multicolumn{2}{c|}{$89.79\%$}& \multicolumn{2}{c|}{$94.61\%$}& \multicolumn{2}{c|}{$100\%$}                   \\ \hline
  0.08$\%$ noise & \multicolumn{2}{c|}{$63.29\%$}       & \multicolumn{2}{c|}{$93.73\%$}& \multicolumn{2}{c|}{$94.03\%$}& \multicolumn{2}{c|}{$100\%$}                   \\ \hline
  0.09$\%$ noise & \multicolumn{2}{c|}{$65.36\%$}       & \multicolumn{2}{c|}{$92.86\%$}& \multicolumn{2}{c|}{$94.81\%$}& \multicolumn{2}{c|}{$100\%$}                   \\ \hline
  0.10$\%$ noise & \multicolumn{2}{c|}{$65.79\%$}       & \multicolumn{2}{c|}{$96.15\%$}& \multicolumn{2}{c|}{$96.15\%$}& \multicolumn{2}{c|}{$100\%$}                   \\ \hline
  0.11$\%$ noise & \multicolumn{2}{c|}{$64.52\%$}       & \multicolumn{2}{c|}{$92.61\%$}& \multicolumn{2}{c|}{$96.00\%$}& \multicolumn{2}{c|}{$100\%$}                   \\ \hline
  0.12$\%$ noise & \multicolumn{2}{c|}{$66.23\%$}       & \multicolumn{2}{c|}{$94.81\%$}& \multicolumn{2}{c|}{$93.75\%$}& \multicolumn{2}{c|}{$100\%$}                   \\ \hline
  0.13$\%$ noise & \multicolumn{2}{c|}{$63.69\%$}       & \multicolumn{2}{c|}{$94.47\%$}& \multicolumn{2}{c|}{$98.79\%$}& \multicolumn{2}{c|}{$100\%$}                   \\ \hline
  0.14$\%$ noise & \multicolumn{2}{c|}{$66.25\%$}       & \multicolumn{2}{c|}{$96.63\%$}& \multicolumn{2}{c|}{$100\%$}& \multicolumn{2}{c|}{$100\%$}                   \\ \hline
  0.15$\%$ noise & \multicolumn{2}{c|}{$63.50\%$}       & \multicolumn{2}{c|}{$91.18\%$}& \multicolumn{2}{c|}{$97.81\%$}& \multicolumn{2}{c|}{$100\%$}                   \\ \hline
  \end{tabular}
  \caption{Precision on different noise scale.}
  \vspace{-1.5em}
\end{table}

\end{document}